\documentclass[aps,prc,reprint,superscriptaddress]{revtex4-2}

\usepackage{graphicx}
\usepackage{amsmath}
\usepackage{float}
\usepackage{subcaption}

\usepackage{xcolor}
    \definecolor{myred}{HTML}{E45F62}
    \definecolor{myblue}{HTML}{5752D9}
    \definecolor{mygreen}{HTML}{318C45}

\usepackage{comment}

\begin{document}


\title{Equilibrated fraction of QCD matter in high-energy oxygen--oxygen collisions}


\author{Naoya Ito}
\email{n-ito-3e5@eagle.sophia.ac.jp}
\affiliation{Department of Physics, Sophia University, Tokyo 102-8554, Japan}

\author{Tetsufumi Hirano}
\email{hirano@sophia.ac.jp}
\affiliation{Department of Physics, Sophia University, Tokyo 102-8554, Japan}


 \date{\today}

\begin{abstract}
    We quantify to what degree the QCD matter created in high-energy oxygen--oxygen ($\mathrm{O}+\mathrm{O}$) collisions at $\sqrt{s_{NN}} = 5.36$ TeV reaches a locally equilibrated state.
For this purpose, we employ a novel framework based on the core--corona picture that describes the dynamics of both locally equilibrated fluids  (the core) and nonequilibrium  particles (the corona).  
Contributions from the core become larger than those from the corona above charged-particle multiplicity at midrapidity, $\langle dN_{\mathrm{ch}}/d\eta\rangle_{|\eta|<0.5} \approx 20$.
We also find that nonnegligible contributions from the corona still remain even in central $\mathrm{O}+\mathrm{O}$ collisions.
The yield ratios of strange baryons to charged pions exhibit an increasing behavior with increasing multiplicity at midrapidity.
However, these ratios are smaller than those obtained when assuming that QCD matter has reached complete chemical equilibrium.
These results demonstrate that a purely hydrodynamic approach is insufficient and that the inclusion of a corona component is essential for describing the dynamics of intermediate-size systems such as  $\mathrm{O}+\mathrm{O}$ collisions.
\end{abstract}


\maketitle


\section{INTRODUCTION}
    \label{sec:introduction}

High-energy nuclear collisions provide a unique opportunity to study the bulk and transport properties of strongly interacting matter, known as the quark gluon plasma (QGP) \cite{Yagi:2005yb}.
The dynamics of the QGP is governed by quantum chromodynamics (QCD).
Collision experiments with various nuclei have been carried out at Relativistic Heavy Ion Collider (RHIC) at Brookhaven National Laboratory (BNL) and Large Hadron Collider (LHC) at CERN.
The QGP has been produced in large collision systems such as Au+Au \cite{BRAHMS:2004adc, PHENIX:2004vcz, PHOBOS:2004zne, STAR:2005gfr} and Pb+Pb \cite{Muller:2012zq} collisions, which have played a central role in establishing its properties such as the equation of state and the shear viscosity.
Signs of collective phenomena have also been observed in small collision systems involving protons, such as $p$ + $p$ and $p$ + Pb collisions \cite{Dusling:2015gta,Song:2017wtw,Nagle:2018nvi,Schenke:2021mxx,Noronha:2024dtq,Grosse-Oetringhaus:2024bwr}.
Therefore, understanding the system-size dependence of QGP production is crucial for clarifying the mechanism of QGP formation.

The strongly interacting QCD matter produced in these collisions undergoes hydrodynamic space–time evolution.
During this evolution, the spatial anisotropy of the QCD matter in the transverse plane immediately after the collision is converted into the momentum anisotropy of the finally observed hadrons \cite{Ollitrault:1992bk,Poskanzer:1998yz,Kolb:2003zi,Hama:2005dz,Alver:2010gr}.
By exploiting this phenomenon, attempts have recently begun to probe the structure of the colliding nuclei themselves, giving rise to a new research direction \cite{Giacalone:2021udy,Jia:2022ozr,STAR:2024wgy,Giacalone:2025vxa}.

Under these circumstances, oxygen–oxygen ($\mathrm{O} + \mathrm{O}$) collision experiments have recently been conducted at RHIC \cite{Huang:2023viw,Zhang:2025cgr} and the LHC \cite{ALICE:2025luc,ATLAS:2025nnt,CMS:2025bta,CMS:2025tga,CMSPre_RapidityDist}.
Oxygen, with a mass number of 16, is a relatively light nucleus that provides an intermediate system size between proton- and heavy-ion collisions.
Therefore, it offers an important opportunity to investigate the system-size dependence of QGP production, particularly in the small-system regime \cite{Brewer:2021kiv}.

This collision system is also of interest from the perspective of nuclear structure.
Oxygen nuclei may exhibit a so-called $\alpha$-cluster structure, in which four $\alpha$ particles (each consisting of two protons and two neutrons) form a clustered configuration \cite{Schuck:2016fex,Wei:2024}.
As mentioned above, the spatial geometry of the colliding nuclei can be reflected in the momentum distributions and correlations of the finally observed hadrons.
Therefore, analyzing these observables in high-energy $\mathrm{O} + \mathrm{O}$ collisions may provide new insights into the structure of oxygen nuclei \cite{Broniowski:2013dia,Zhang:2017xda,Rybczynski:2017nrx,Lim:2018huo,Rybczynski:2019adt,Li:2020vrg,Summerfield:2021oex,Behera:2021zhi,YuanyuanWang:2024sgp,Giacalone:2024luz,Zhang:2024vkh,Shafi:2025feq,Hu:2025eid,Loizides:2025ule,Constantin:2025ova}.

In high-energy $\mathrm{O} + \mathrm{O}$ collisions, a key issue in connecting the system-size dependence of QGP production with the structure of the colliding nuclei is the degree to which the produced system achieves local equilibrium shortly after the collision.
If local equilibrium is reached, the subsequent space–time evolution of the system can be described by using relativistic hydrodynamics.
In this case, the spatial anisotropy of the system immediately after the collision is efficiently converted into the momentum anisotropy of the hadrons observed in the final state.
On the other hand, if the system reaches only partial local equilibration, it is not obvious how this relationship is quantitatively realized.
To describe the partial local equilibration, the core--corona picture was proposed to explain the centrality dependence of strange hadron yield ratios \cite{Werner:2007bf} and has been implemented into the Monte Carlo event generator \cite{Werner:2005jf,Werner:2013tya,Pierog:2013ria,Werner:2023mod}. 

In this paper, we quantify the fraction of hadrons originating from locally equilibrated QCD matter in high-energy $\mathrm{O} + \mathrm{O}$ collisions.
To this end, we employ a dynamical core–corona initialization (DCCI) model \cite{Kanakubo:2019ogh, Kanakubo:2021qcw,Kanakubo:2022ual}.
In this model, locally equilibrated matter (the core) is dynamically generated from nonequilibrium partons produced immediately after the collision, and its space--time evolution is described by relativistic hydrodynamics.
By simultaneously describing the evolution of matter that does not reach local equilibrium (the corona), the entire collision system is treated within a two-component framework.
Hadrons produced from locally equilibrated matter are referred to as the core component, whereas hadrons produced from nonequilibrium partons via string fragmentation constitute the corona component.
In the DCCI model, these two components can mix through final-state hadron--hadron interactions treated by using a hadron cascade model.
However, since these interactions have only a minor effect on the hadron yields, we can determine the fraction of hadrons originating from locally equilibrated matter by switching off the final-state hadron--hadron interactions.

The present paper is organized as follows: We briefly explain the DCCI model in Sec.~\ref{sec:model}.
In Sec.~\ref{sec:results}, we show the results from the
DCCI model in $\mathrm{O} + \mathrm{O}$ collisions at the LHC energy.
We discuss the multiplicity and transverse momentum dependences of the fractions of the core and the corona components and analyze particle yield ratios of strange baryons to charged pions at midrapidty.
Section IV is devoted to the summary of the present paper.

Throughout this paper, we use the natural
units, $\hbar$ = $c$ = $k_{B}$ = $1$.
    
\section{MODEL}
    \label{sec:model}

In this paper, we employ the Dynamical Core--Corona Initialization (DCCI) framework \cite{Kanakubo:2019ogh, Kanakubo:2021qcw,Kanakubo:2022ual} to describe the dynamics of $\mathrm{O} + \mathrm{O}$ collisions at $\sqrt{s_{NN}} = 5.36$ TeV. 
We specifically use version 2 of the DCCI model (hereafter denoted as DCCI2) in which several crucial updates have been made \cite{Kanakubo:2021qcw,Kanakubo:2022ual} from version 1 \cite{Kanakubo:2019ogh}.
These include a more sophisticated way to treat four-momentum deposition of initial partons and Monte Carlo sampling of particlization of the fluid elements on the switching hypersurface.
The DCCI2 framework provides a comprehensive and dynamical description of high-energy nuclear collisions by simultaneously treating the equilibrated matter (predominantly the thermalized QGP) as core components and the nonequilibrated particles (predominantly high-momentum partons) as corona components. 
In this section, we briefly introduce the main ingredients of the DCCI2 framework and explain some changes from the original framework for the purpose to analyze $\mathrm{O} + \mathrm{O}$ collisions. 
For a complete and detailed description of the DCCI2 framework, we refer the reader to Refs.~\cite{Kanakubo:2021qcw,Kanakubo:2022jju}.

In the DCCI2 framework, initial partons are produced by using the event generator PYTHIA8 Angantyr~\cite{PYTHIA,Bierlich:2018xfw}.
In PYTHIA, we employ the GLISSANDO 2~\cite{Rybczynski:2013yba} model to randomly generate the initial configuration of nucleons in the colliding oxygen nuclei.
This is the default model for nuclear geometry in heavy ion mode in PYTHIA \footnote{Although the GLISSANDO 2 model employs a Woods–Saxon nuclear density profile and is therefore primarily optimized for heavier nuclei, it is adopted here for consistency. The detailed modeling of nuclear geometry has recently become an active topic in high-energy nuclear collisions. In the present work, our aim is to establish a controlled baseline for future systematic studies of geometry effects. For this purpose, we retain the same modeling framework across different systems, including light nuclei, to ensure internal consistency and a transparent comparison.}.
The initial partons are produced as a superposition of multiple nucleon--nucleon sub-collisions in the Angantyr model. 
In this analysis, we have replaced PYTHIA 8.244, originally employed in the DCCI2 model \cite{Kanakubo:2021qcw}, with PYTHIA 8.315.
This replacement is a crucial first step for our future research. The latest versions of PYTHIA offer expanded options for selecting various nuclear density profiles, thereby establishing a baseline for future systematic investigations of nuclear profile effects.
We have verified that the effects of this update on the particle yields are negligible.

We basically use the default parameter set in PYTHIA8 Angantyr except for two parameters, \texttt{MultipartonInteraction:pT0Ref} and \texttt{SpaceShower:pT0Ref}. These parameters control the yields of initial partons by regularizing the cross section of multiparton interactions and infrared QCD emission \cite{Sjostrand:2017cdm}.
In the previous analysis of $\mathrm{Pb} + \mathrm{Pb}$ collisions at $\sqrt{s_{NN}} = 2.76$ TeV,  $p_{\mathrm{T0Ref}}= 1.0$ GeV was used for both parameters. In this paper, we also use the same value for $p_{\mathrm{T0Ref}}$ to calculate observables in $\mathrm{O} + \mathrm{O}$ collisions.

The produced nonequilibrium partons dynamically deposit their energy and momentum into the medium according to a prescribed four-momentum deposition rate \cite{Kanakubo:2021qcw}.
The magnitude of this rate strongly depends on the local density of surrounding partons, thereby encoding the core–corona structure of the system.
Here, the surrounding partons include not only those produced in the initial stage but also those already incorporated into the equilibrated medium.
These energy–momentum depositions act as source terms in the (3+1)-dimensional ideal hydrodynamic equations \cite{Okai:2017ofp}, which describe the formation and subsequent fluid-dynamical evolution of the equilibrated matter identified as the core component.
Meanwhile, nonequilibrium partons that do not fully lose their energy and momentum continue to propagate and are classified as the corona component.

When the local temperature of the fluid element drops below the specific switching temperature $T_{\text{sw}} = 165$ MeV, the core component is converted into hadrons via the Monte Carlo sampler iS3D~\cite{McNelis:2019auj} based on the Cooper-Frye prescription \cite{Cooper:1974mv}.
In parallel, nonequilibrated partons that did not reach thermal equilibrium undergo hadronization through the Lund string fragmentation in PYTHIA. For this fragmentation process, we continue to employ PYTHIA 8.244 to ensure consistency with our established hadronization framework, even though the initial parton production was updated to a newer version of PYTHIA.


Finally, hadrons originating from both the core and corona components are passed to the hadronic transport model JAM~\cite{JAM} to account only for resonance decays in this paper.
Although the JAM module within the DCCI2 framework generally includes both hadronic rescatterings and resonance decays, the results presented in this paper are obtained with hadronic rescatterings switched off in order to isolate the fractional contributions of the core and corona components.
This is because hadronic rescatterings mix up these two
components in the late stage of reactions.
Our numerical studies confirm that hadronic rescatterings have a negligible impact on the final rapidity distributions and on the integrated particle yields in this collision system.
Therefore, disabling hadronic rescatterings enables a clean separation of the core and corona contributions without affecting the primary observables.

\section{RESULT}
    \label{sec:results}

We generate $10^5$ minimum-bias events for $\mathrm{O} + \mathrm{O}$ collisions at $\sqrt{s_{NN}} = 5.36$ TeV using the DCCI2 model introduced in Sec.~\ref{sec:model}.
As discussed in the previous section, hadronic rescatterings mix the contributions from the core and corona components.
We have verified, however, that the final hadron yields are not significantly modified by hadronic rescatterings. We therefore neglect their effects on the final hadron yields in the following analysis.
All results presented in this section are obtained with hadronic rescatterings switched off and hadron decays switched on, allowing us to focus on the relative contributions of the core and corona components.

In our analysis using PYTHIA in heavy-ion mode, individual events are not necessarily generated with unit weight. Therefore, event averages must be computed as weighted averages rather than simple arithmetic means. For an observable $O$, the event-averaged value is evaluated as
\begin{equation}
\langle O \rangle = \frac{\sum_i w_i O_i}{\sum_i w_i},
\end{equation}
where $w_i$ denotes the event weight provided by PYTHIA and $O_i$ is the value of the observable in the $i$th event. Here the weight $w_i$ is obtained from \texttt{pythia.info.weight()} in PYTHIA. 
All histograms and derived quantities are filled using these event weights, and the final normalization is performed with the sum of weights. This procedure ensures a consistent treatment of the effective cross section and the impact-parameter sampling implemented in the heavy-ion configuration.

When applying a centrality cut, the same weighted procedure is adopted within the selected event class. Namely, events are first classified according to either transverse energy in the range $3<|\eta|< 5$ (Secs.~\ref{sec:charged-hadron-yields} and \ref{sec:core-corona-fraction}) or charged particle multiplicity in the range $2.8 < \eta < 5.1$ and $-3.7 < \eta < -1.7$ (Sec.~\ref{sec:strangeness-enhancement}), depending on the detector coverage to measure the observables. 
Then, only events belonging to the chosen centrality interval are included in the sums. The centrality-dependent average is then computed as
\begin{equation}
\langle O \rangle_{\mathrm{cent}} = \frac{\sum_{i\in \mathrm{cent}} w_i O_i}{\sum_{i\in \mathrm{cent}} w_i},
\end{equation}
where the summation runs over events satisfying the centrality selection. This guarantees that the centrality-dependent observables consistently reflect the weighted event sample and the underlying impact-parameter distribution.

\subsection{Pseudorapidity and centrality dependences of charged hadron yields}
\label{sec:charged-hadron-yields}

\begin{figure}[htbp]
    \centering

    \includegraphics[width=0.97\linewidth, page=1]{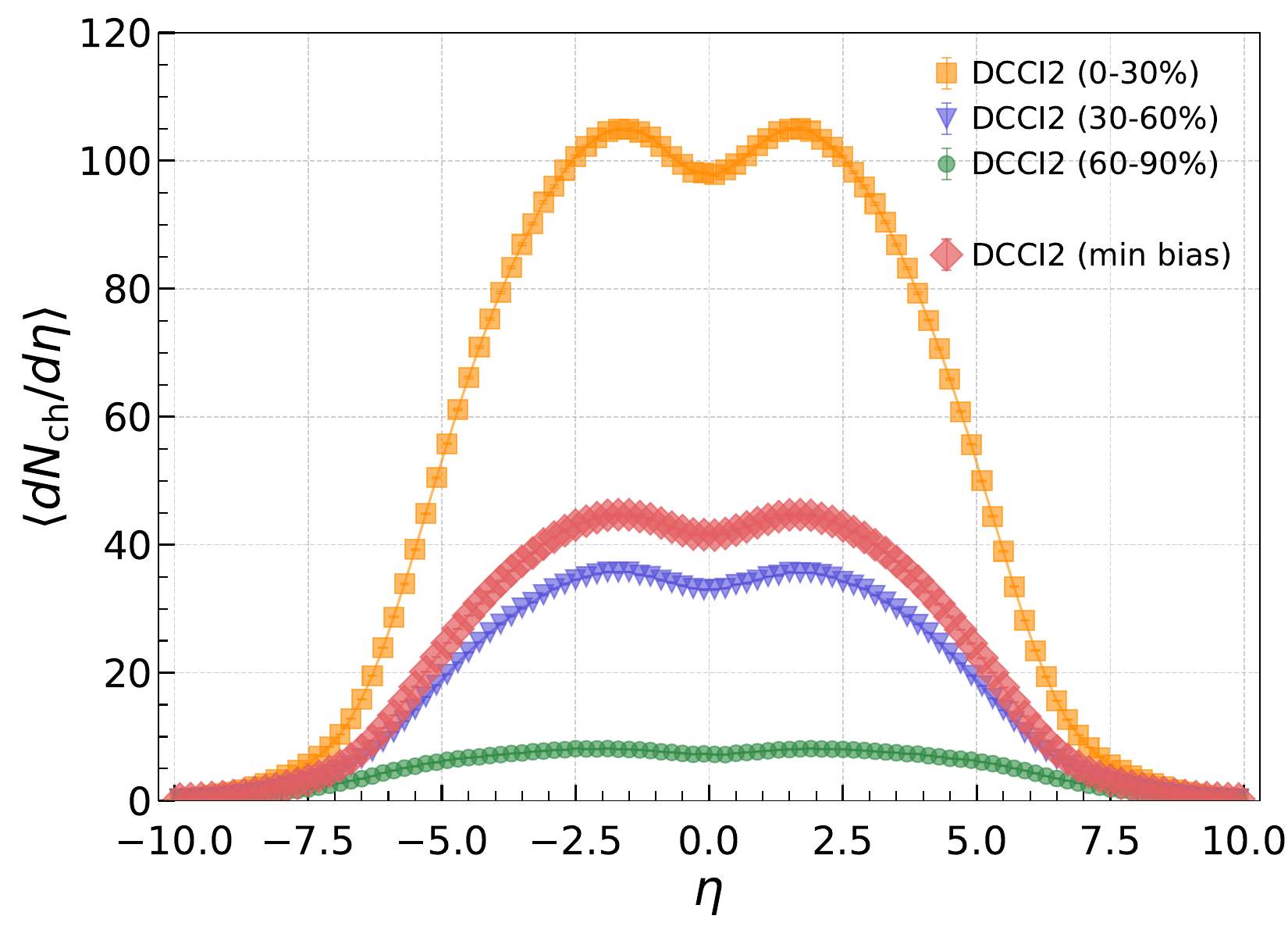}
    
    \caption{Pseudorapidity distribution of charged hadrons, $dN_{\mathrm{ch}}/d\eta$, in $\mathrm{O}+\mathrm{O}$ collisions at $\sqrt{s_{NN}}=5.36$ TeV from the DCCI2 model. 
    From top to bottom, results at 0--30\%, 0--100\%, 30--60\%, and 60--90\% centrality classes are shown for comparison.
    }
    \label{Fig:PseudoRapidityDistribution}
\end{figure}

Figure~\ref{Fig:PseudoRapidityDistribution} shows the pseudorapidity distributions of charged hadrons, $\langle dN_{\mathrm{ch}}/d\eta \rangle$, in $\mathrm{O}+\mathrm{O}$ collisions at $\sqrt{s_{NN}} = 5.36$ TeV from the DCCI2 model.
The result at 0-100\% centrality (minimum bias) is comparable with the preliminary result of $\langle dN_{\mathrm{ch}}/d\eta \rangle \approx 40$ in $|\eta| < 2.4$ from CMS Collaboration~\cite{CMSPre_RapidityDist}.
As discussed in Sec.~\ref{sec:model}, we employed the same parameter set in DCCI2 model as the one in $\mathrm{Pb}+\mathrm{Pb}$ collisions, except for the species of colliding nuclei in PYTHIA. This means that the DCCI2 model well describes the final yields from intermediate to large colliding systems at LHC energies in a unified way.
The results at 0--30\%, 30--60\%, and 60--90\% centrality are also shown for comparison.

\begin{figure}[htbp]
    
    \includegraphics[width=0.8\linewidth, page=1]{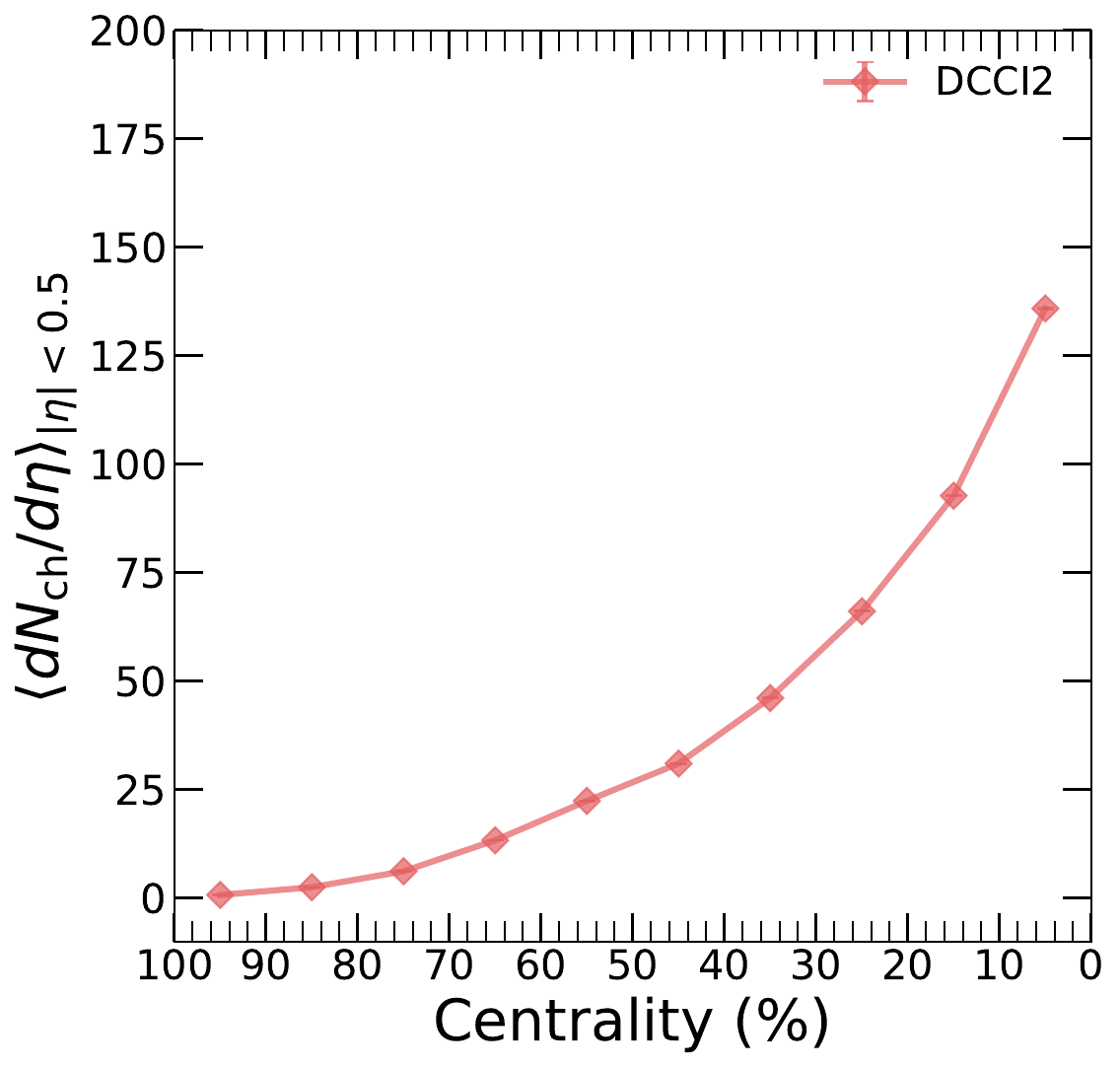}

    \caption{Charged-particle multiplicity at midrapidity $\langle dN_{\mathrm{ch}}/d\eta \rangle _{|\eta|<0.5}$
    for a given centrality. 
    Data points are shown for each 10\% centrality.
    }
    \label{Fig:CentralityDependenceofMultiplicity}
\end{figure}

To see more details on particle production in $\mathrm{O}+\mathrm{O}$ collisions, we show the 
centrality dependence of charged-particle multiplicity at midrapidity $\langle dN_{\mathrm{ch}}/d\eta \rangle _{|\eta|<0.5}$ in Fig.~\ref{Fig:CentralityDependenceofMultiplicity}.
This is again comparable with the preliminary CMS data~\cite{CMSPre_RapidityDist}.
The charged hadron yield at midrapidity reaches $\approx 136$ at 0--10\% centrality. This value is almost identical to the result at 60--70\% centrality in $\mathrm{Pb}+\mathrm{Pb}$ collisions at $\sqrt{s_{NN}}= 5.02$ TeV.
In the previous study \cite{Kanakubo:2021qcw}, it was found that the core contribution exceeds that of the corona for $dN_{\text{ch}}/d\eta \gtrsim 20$. 
This suggests that $\mathrm{O}+\mathrm{O}$ collisions at $\sqrt{s_{NN}} = 5.36$ TeV provide an opportunity to investigate the transition from core-dominated to corona-dominated particle production in the 50--70\% centrality class.
We will discuss this in the following subsections.

\subsection{Fraction of core and corona components}
\label{sec:core-corona-fraction}

\begin{figure}[htbp]
    \centering
    \includegraphics[width=0.9\linewidth, page=1]{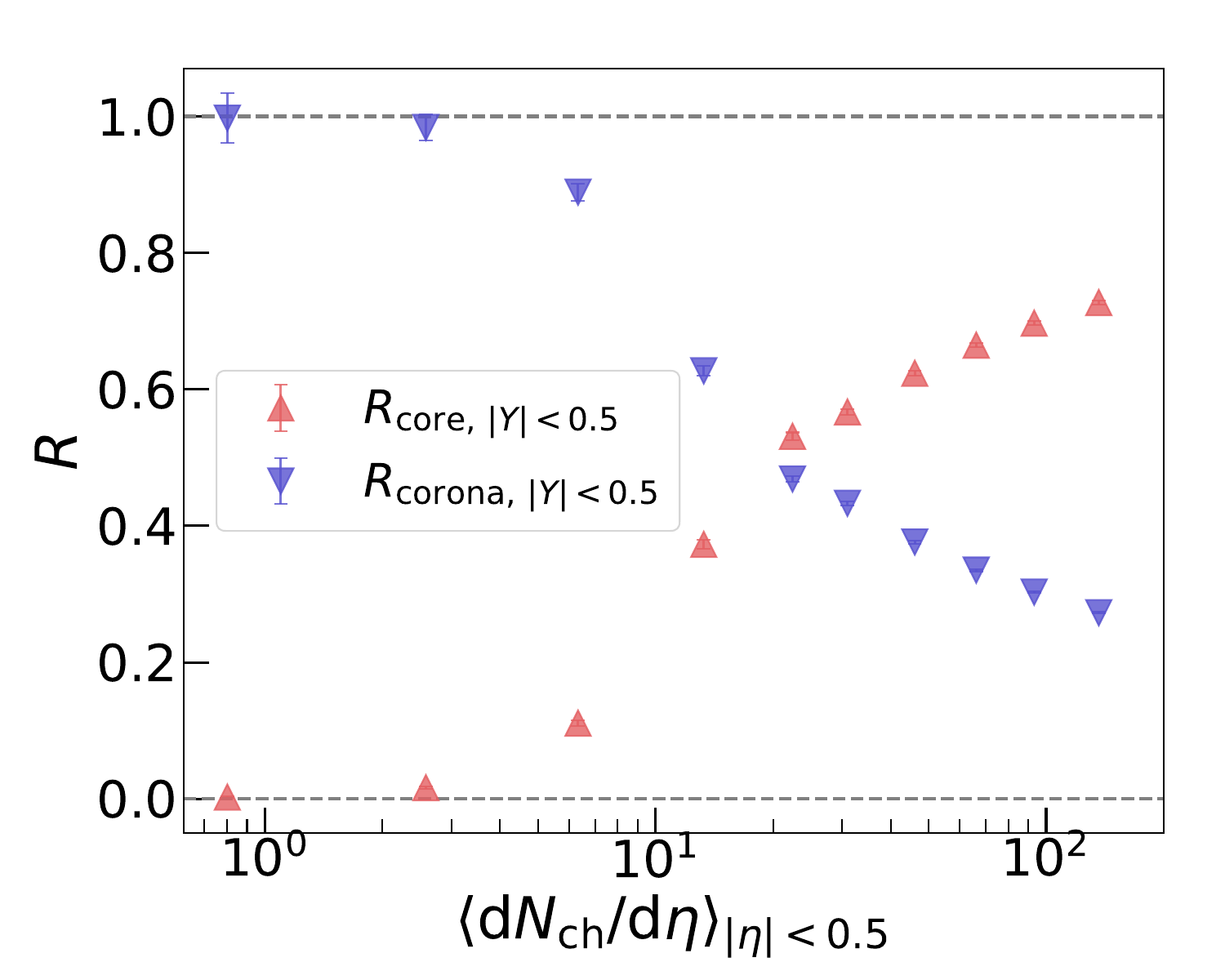}
    \caption{The fractions of the core (red squares) and corona (blue diamonds) components in the hadron yields in $|Y|<0.5$ as functions of charged-particle multiplicity at midrapidity ($|\eta|<0.5$). 
    Data points are shown for each 10\% centrality. 
    }
    \label{Fig:CoreCoronaFraction}
\end{figure}

Figure~\ref{Fig:CoreCoronaFraction} shows the fractions of the core and  corona components of the charged hadrons directly from the switching hypersurface and from resonance decays as functions of multiplicity at midrapidity ($|\eta|<0.5$). 
The fractions of the core and corona components are obtained as
 \begin{equation}
     R_{\mathrm{core/corona}}= \frac{\langle dN_{\mathrm{ch}}/dY \rangle_{\mathrm{core/corona},\ |Y|<0.5}}{\langle dN_{\mathrm{ch}}/dY \rangle_{\mathrm{total},\ |Y|<0.5}}
 \end{equation}
for each centrality. Here $Y$ is rapidity of each hadron.
As shown in Fig.~\ref{Fig:CoreCoronaFraction}, the corona contribution dominates in the low-multiplicity region corresponding to the 80–100\% centrality class.
As the multiplicity increases, the corona fraction decreases and crosses the core fraction at
$\langle dN_{\mathrm{ch}}/d\eta \rangle_{|\eta|<0.5} \approx 20$.
This crossing point may be interpreted as the onset of locally equilibrated matter production.

The results shown in Fig.~\ref{Fig:CoreCoronaFraction} also demonstrate that the corona contribution remains nonnegligible across all centrality classes. Even in the most central (0--10\%) $\mathrm{O} + \mathrm{O}$ collisions, the corona component still accounts for approximately 30\% of the total hadron yield.
Because the system size in $\mathrm{O} + \mathrm{O}$ collisions is insufficient to reach very high multiplicities, unlike in $\mathrm{Pb} + \mathrm{Pb}$ collisions, the corona contribution plays an essential role in understanding the dynamics of such intermediate-sized collision systems.

\begin{figure*}[ht]
    \begin{subfigure}[b]{0.328\linewidth}
        \centering
        \includegraphics[width=\linewidth, page=1]{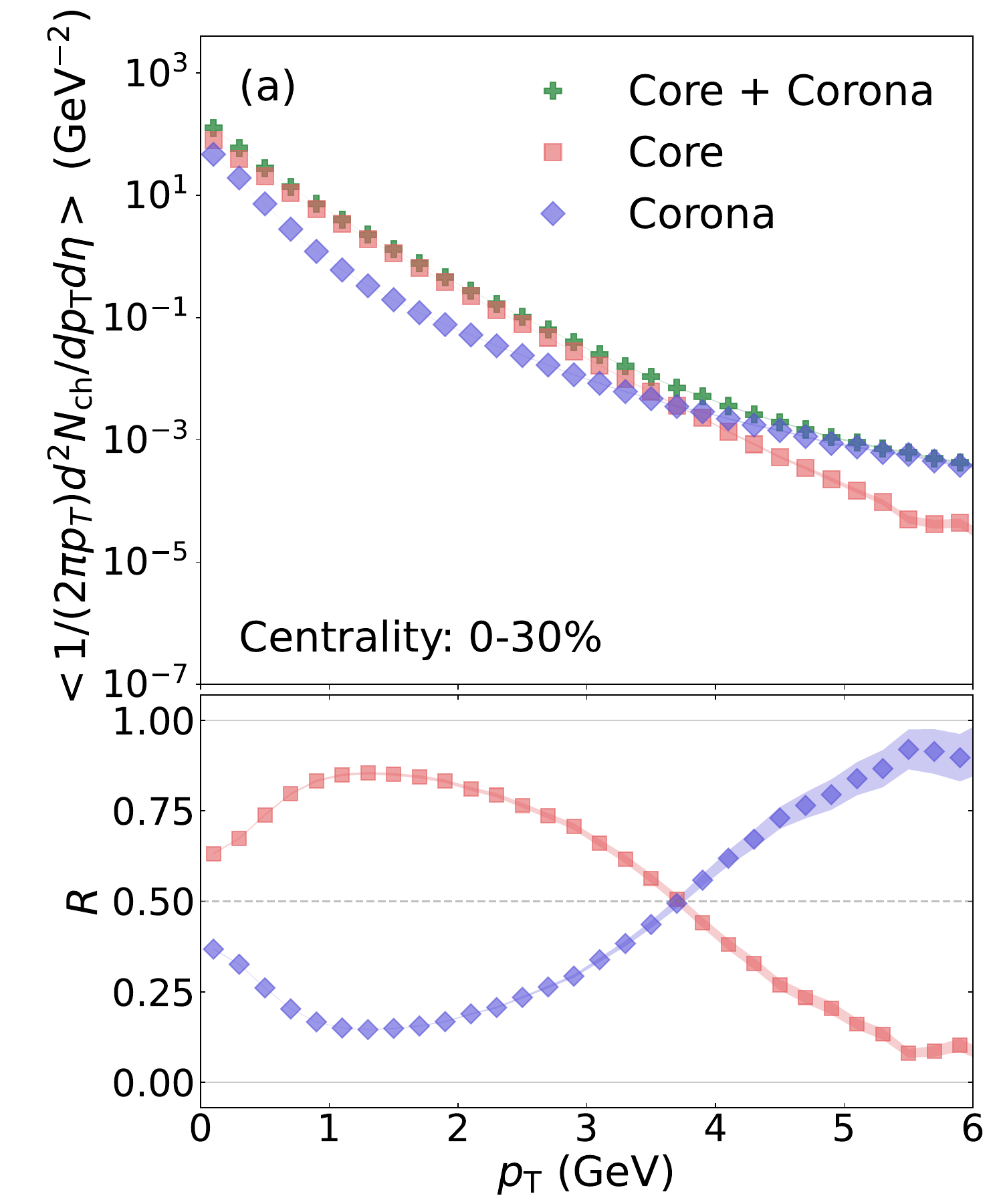}
        \label{Fig:sub_a}
    \end{subfigure}
    \begin{subfigure}[b]{0.328\linewidth}
        \centering
        \includegraphics[width=\linewidth, page=1]{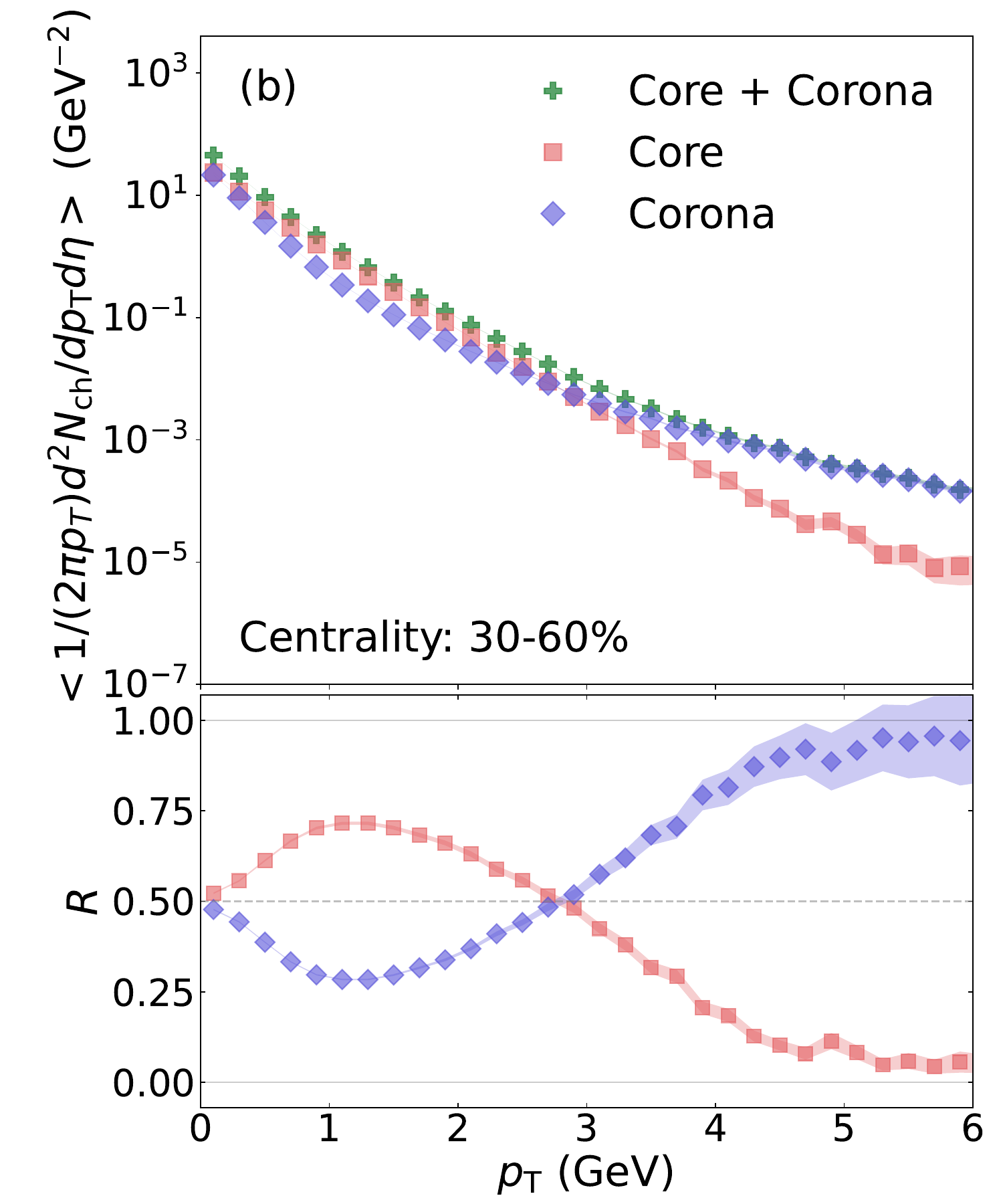}
        \label{Fig:sub_b}
    \end{subfigure}
    \begin{subfigure}[b]{0.328\linewidth}
        \centering
        \includegraphics[width=\linewidth, page=1]{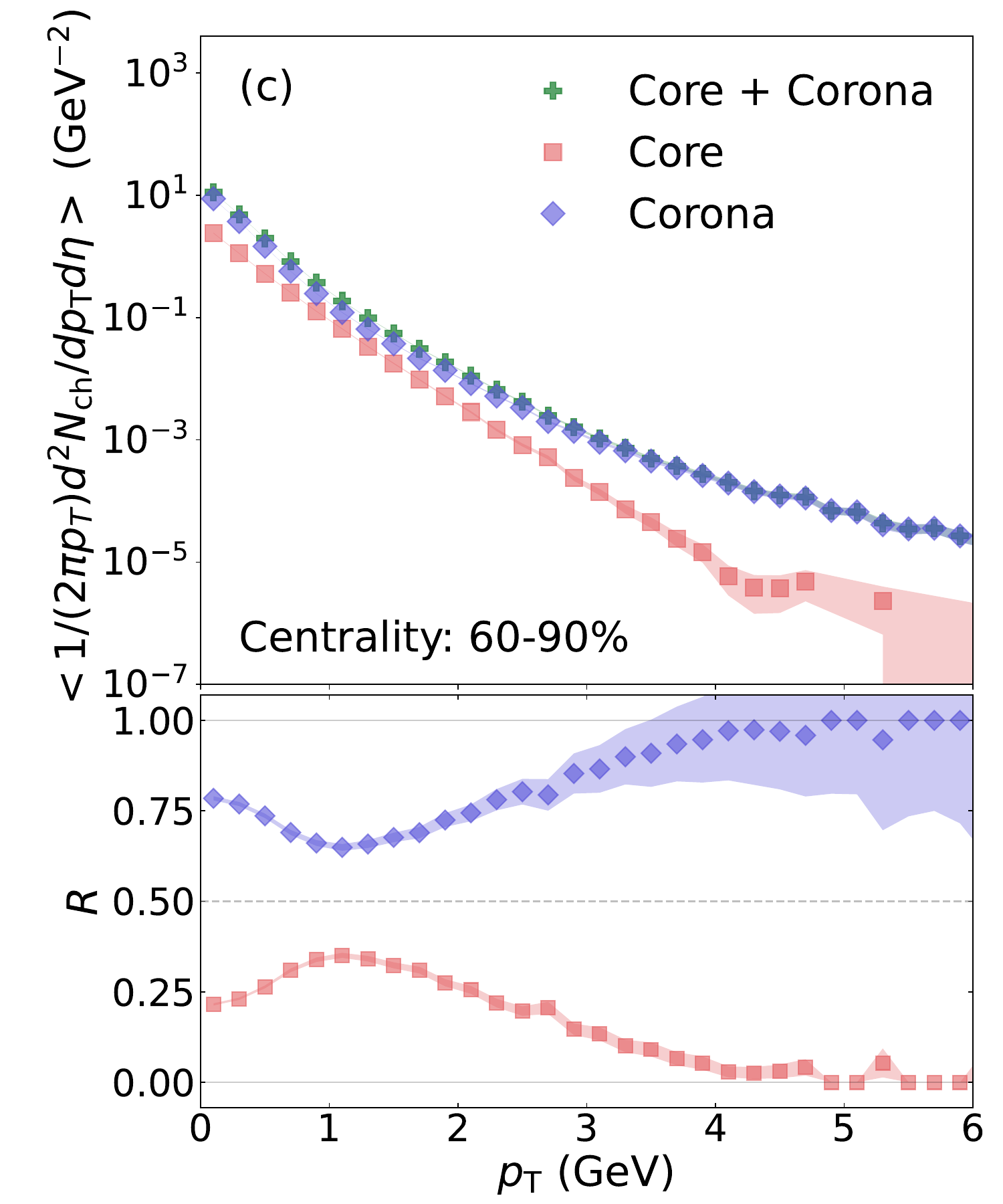}
        \label{Fig:sub_c}
    \end{subfigure}
    \vspace{-15pt}
    \caption{
    Transverse momentum ($p_T$) spectra of charged particles in $p_T > 0.2$~GeV and $|\eta| < 2.4$ in $\mathrm{O}+\mathrm{O}$ collisions at (a) 0--30\%, (b) 30--60\%, and (c) 60--90\% centrality classes from the DCCI2 model.
    Shaded bands represent the statistical error.
    Upper panels: Total yields (green crosses) and their breakdown into the core (red squares) and the corona (blue diamonds) contributions.
    Lower panels: Corresponding fractions of the core component (red squares) and corona component (blue diamonds). 
}
    \label{Fig:CoreCoronapTDist_Total}
\end{figure*}

The upper panels of Fig.~\ref{Fig:CoreCoronapTDist_Total} show the $p_T$ distributions of charged hadrons and their individual contributions from the core and corona components in $\mathrm{O}+\mathrm{O}$ collisions at (a) 0--30\%, (b) 30--60\%, and (c) 60--90\% centrality from the DCCI2 model.
On the other hand, the lower panels show the corresponding fractional contributions, $R_{\mathrm{core}}$ and $R_{\mathrm{corona}}$, as functions of $p_{T}$, where $R_{\mathrm{core}/\mathrm{corona}}$ are calculated as
\begin{equation}
    R_{\mathrm{core/corona}}(p_T) = 
    \frac{
    \left \langle \frac{1}{2 \pi } \frac{dN^2_{\mathrm{ch}}}{p_Tdp_Td\eta} \right\rangle_{\mathrm{core/corona}}
    }
    {
    \left\langle \frac{1}{2 \pi} \frac{dN^2_{\mathrm{ch}}}{ p_Tdp_Td\eta} \right \rangle_{\mathrm{total}}
    }\,.
    \end{equation}

These results clearly indicate that the fraction from the core (the corona) decreases (increases) above $p_{T} \approx 1.5$ GeV.
This is qualitatively consistent with the known fact in high-energy nuclear collisions that hadrons in the low $p_T$ region  originate from equilibrated matter, while those in the high $p_T$ region are from the fragmentation of hard nonequilibrated partons.
The competition between the core and the corona components results from the interplay between different shapes of the spectra: an exponential function for Lorentz-boosted equilibrated distributions and a power-law function for the fragmentation of nonequilibrated partons.
We quantify the scale of $p_T$ at which the dominant contribution changes from the core to the corona.
As $p_T$ increases, the dominant contribution changes from the core to the corona at $p_T \approx 3.7$ GeV at 0--30\% centrality.
On the other hand, it reduces in more peripheral collisions: it crosses at $p_T \approx 2.8$ GeV at 30--60\% centrality, and the contribution from the core components never dominates in the entire $p_T$ region at 60--90\% centrality.
This observation is consistent with the trend established in Fig.~\ref{Fig:CoreCoronaFraction}, where the relative contribution of the corona component becomes more prominent as one moves to more peripheral collisions.

As shown in Fig.~\ref{Fig:CoreCoronapTDist_Total}, we observe a substantial corona contribution even at low transverse momentum, $p_{T} \lesssim 1$ GeV, consistent with the findings of Ref.~\cite{Kanakubo:2022ual}.
This “soft-from-corona” component was previously identified as a source of dilution of the four-particle cumulant $c_2\{4\}$, which would otherwise reflect purely core-driven correlations \cite{Kanakubo:2022ual}.
These results highlight the essential role of the corona contribution in the analysis of bulk ($p_T$-integrated) observables and indicate that incorporating the soft component of corona origin is necessary for a comprehensive description of hadron production, even in the very low-$p_T$ region.

\begin{figure*}
    \centering
    \begin{subfigure}[b]{0.49\linewidth}
        \centering
        \includegraphics[width=0.7\linewidth]{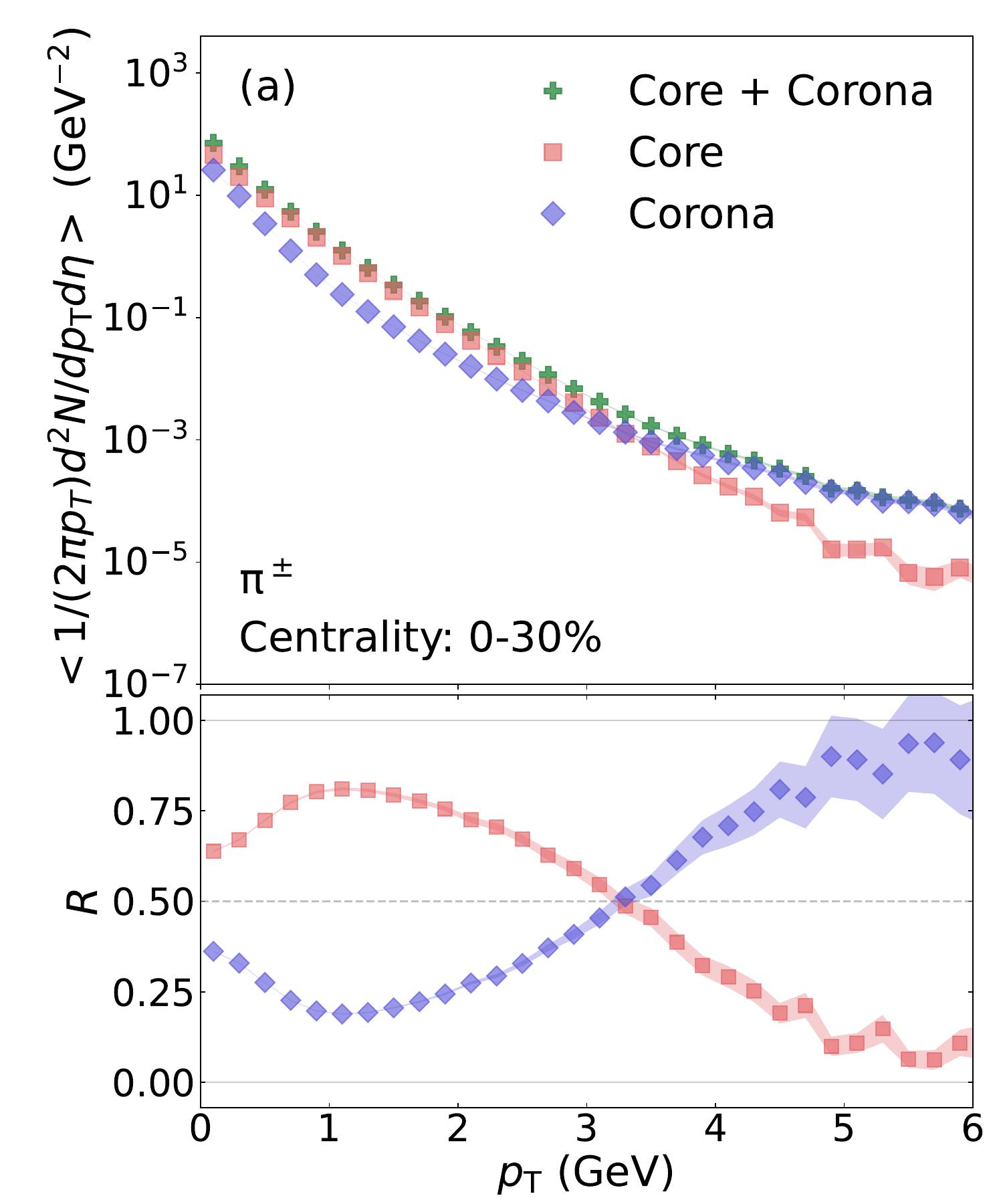}
    \end{subfigure}
    \hfill
    \begin{subfigure}[b]{0.49\linewidth}
        \centering
        \includegraphics[width=0.7\linewidth]{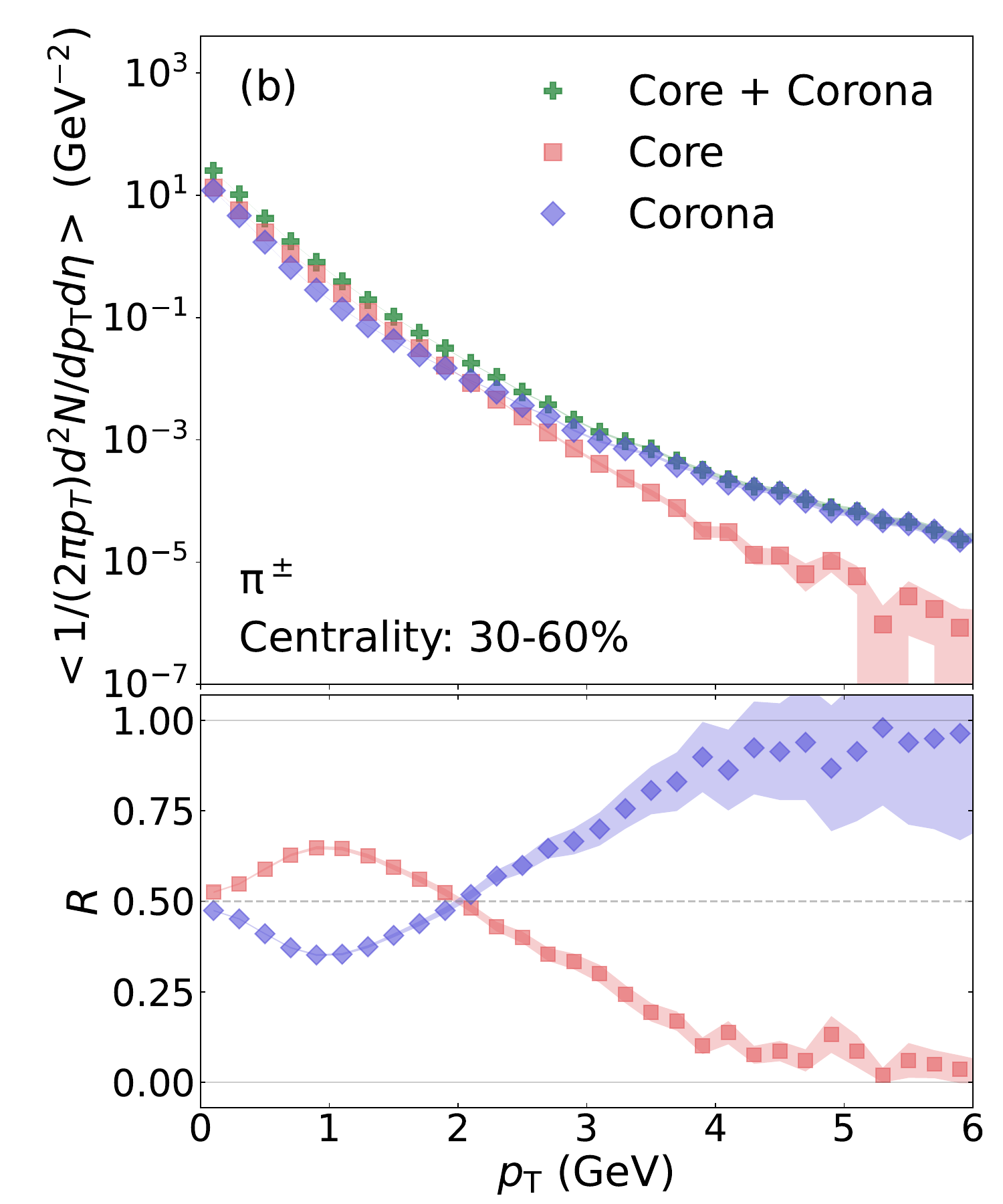}
    \end{subfigure}


    \begin{subfigure}[b]{0.49\linewidth}
        \centering
        \includegraphics[width=0.7\linewidth]{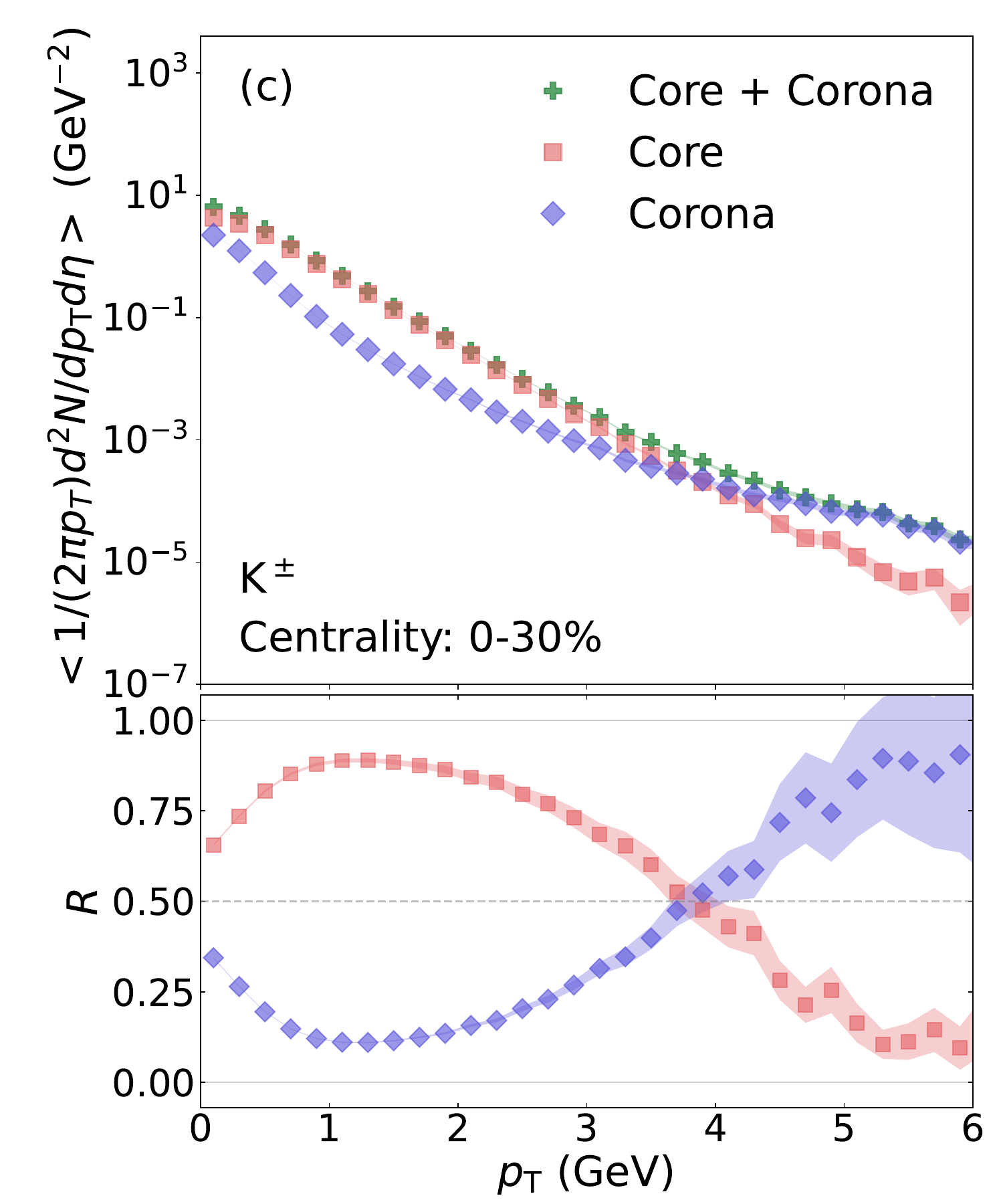}
    \end{subfigure}
    \hfill
    \begin{subfigure}[b]{0.49\linewidth}
        \centering
        \includegraphics[width=0.7\linewidth]{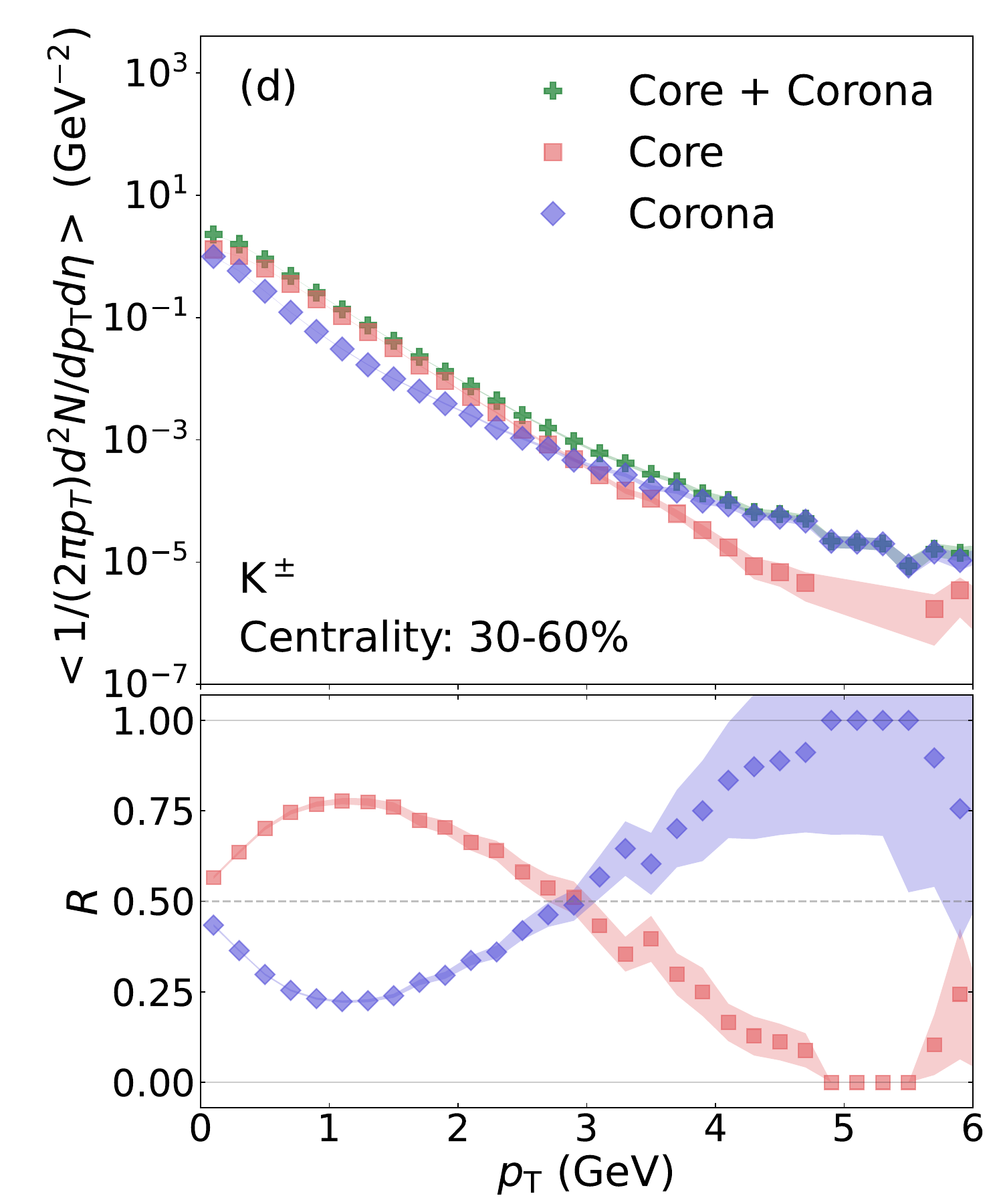}
    \end{subfigure}


    \begin{subfigure}[b]{0.49\linewidth}
        \centering
        \includegraphics[width=0.7\linewidth]{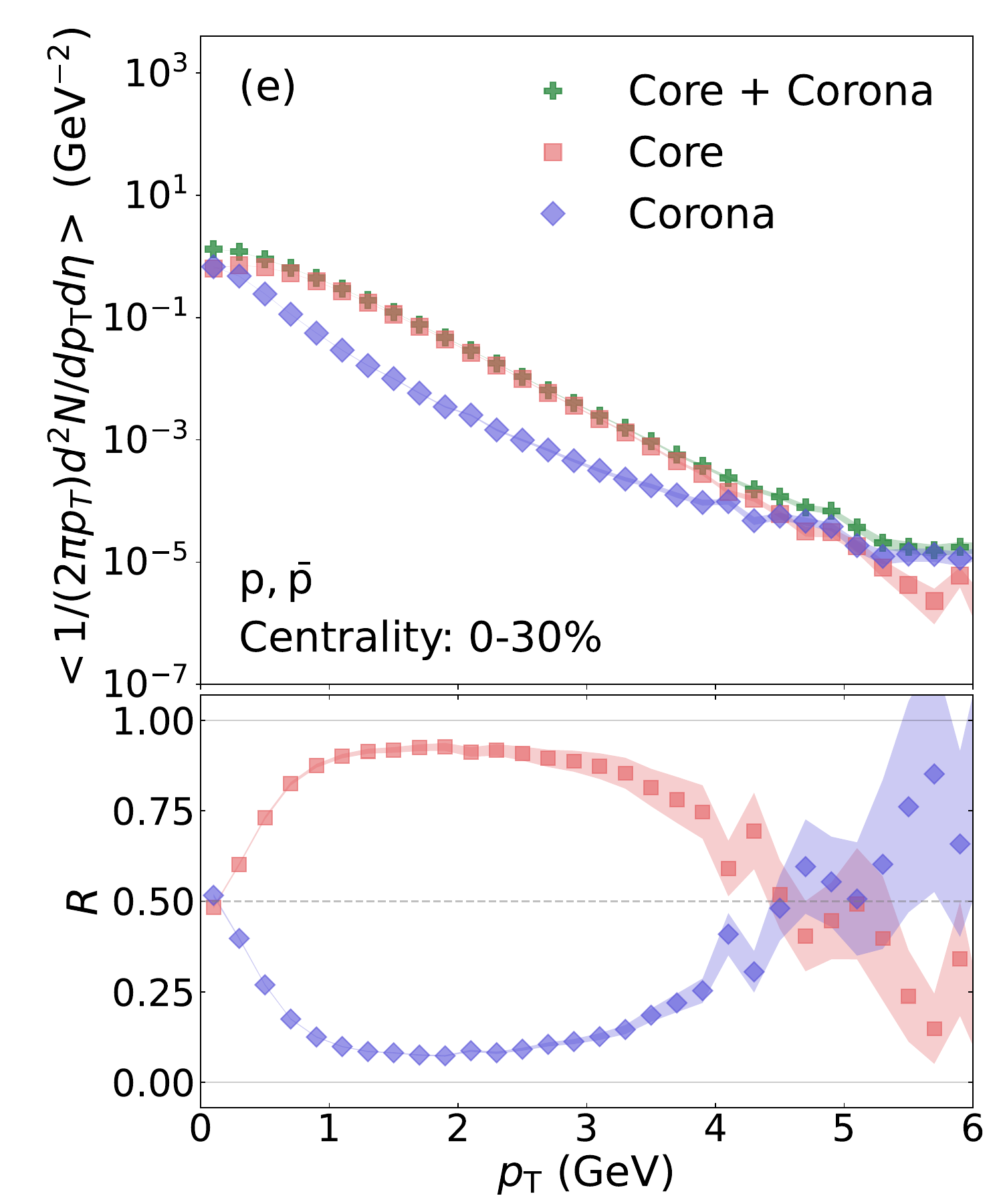}
    \end{subfigure}
    \hfill
    \begin{subfigure}[b]{0.49\linewidth}
        \centering
        \includegraphics[width=0.7\linewidth]{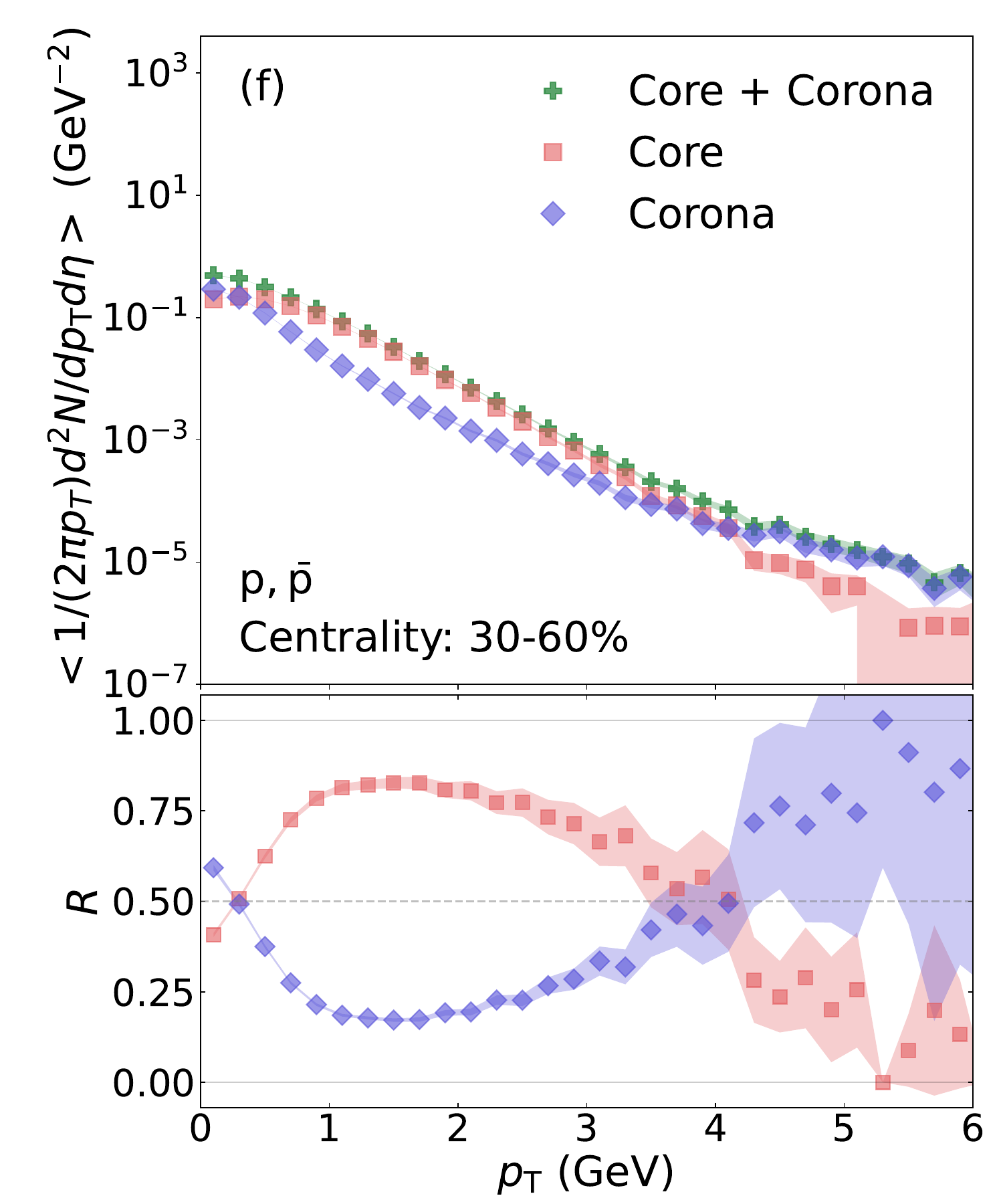}
    \end{subfigure}

    \caption{
     Transverse momentum  spectra of charged pions, charged kaons, and protons and antiprotons in $p_T > 0.2$~GeV and $|\eta| < 2.4$ in $\mathrm{O}+\mathrm{O}$ collisions from the DCCI2 model.
     Results are shown for (a) $\pi^\pm$ at 0--30\%, (b) $\pi^\pm$ at 30--60\%, (c) $K^\pm$ at 0--30\%, 
    (d) $K^\pm$ at 30--60\%, 
    (e) $p$ and $\bar{p}$ at 0--30\%, and 
    (f) $p$ and $\bar{p}$ at 30--60\% centrality classes.
    Shaded bands represent the statistical error.
    Upper panels: Total yields (green crosses) and their breakdown into the core (red squares) and the corona (blue diamonds) contributions.
    Lower panels: Corresponding fractions of the core component (red squares) and corona component (blue diamonds).  
    }
    \label{Fig:CoreCoronapTDist_pid}
\end{figure*}

Figure \ref{Fig:CoreCoronapTDist_pid} shows transverse momentum spectra of charged pions, charged kaons, and protons and antiprotons in $\mathrm{O}+\mathrm{O}$ collisions from the DCCI2 model. Left (Right) panels in Fig.~\ref{Fig:CoreCoronapTDist_pid} exhibit results at 0--30\% (30--60\%) centrality class. 
The overall tendencies are qualitatively similar to the ones at the corresponding centrality class shown in Fig.~\ref{Fig:CoreCoronapTDist_Total}: The core dominance in the intermediate $p_T$ region, the corona dominance in the high $p_T$ region, and competition between these two contributions in the very low $p_T$ region.

For all particle species, the crossing point, at which the contributions of the core and corona components intersect, shifts toward lower $p_T$ as the collision becomes more peripheral. 
A more equilibrated medium is formed in more central collisions, making the core component dominant up to a higher $p_T$ range.

When comparing particle species, it is clear that the onset momentum moves toward higher $p_T$ as the particle mass increases \cite{Hirano:2003pw,Kanakubo:2022ual}. 
Specifically for protons, the region where the core component is dominant extends to higher $p_T$ ($p_T \lesssim 5$ GeV) compared to pions ($p_T \lesssim 3.3$ GeV) and kaons ($p_T \lesssim 3.8$ GeV) at 0--30\% centrality. 
This strongly reflects the effect of radial flow in the core component.
During the hydrodynamic expansion, heavier particles receive a larger boost from the collective flow and are pushed toward higher $p_T$. 
As a result, the proton distribution remains dominated by the core component even at higher momentum. These results support the fact that the core evolves as a dynamic medium with collective flow even in $\mathrm{O}+\mathrm{O}$ collisions.

\subsection{Strangeness hadron yields}
\label{sec:strangeness-enhancement}

\begin{figure}[htbp]
    \centering
    \includegraphics[width=0.8\linewidth, page=1]{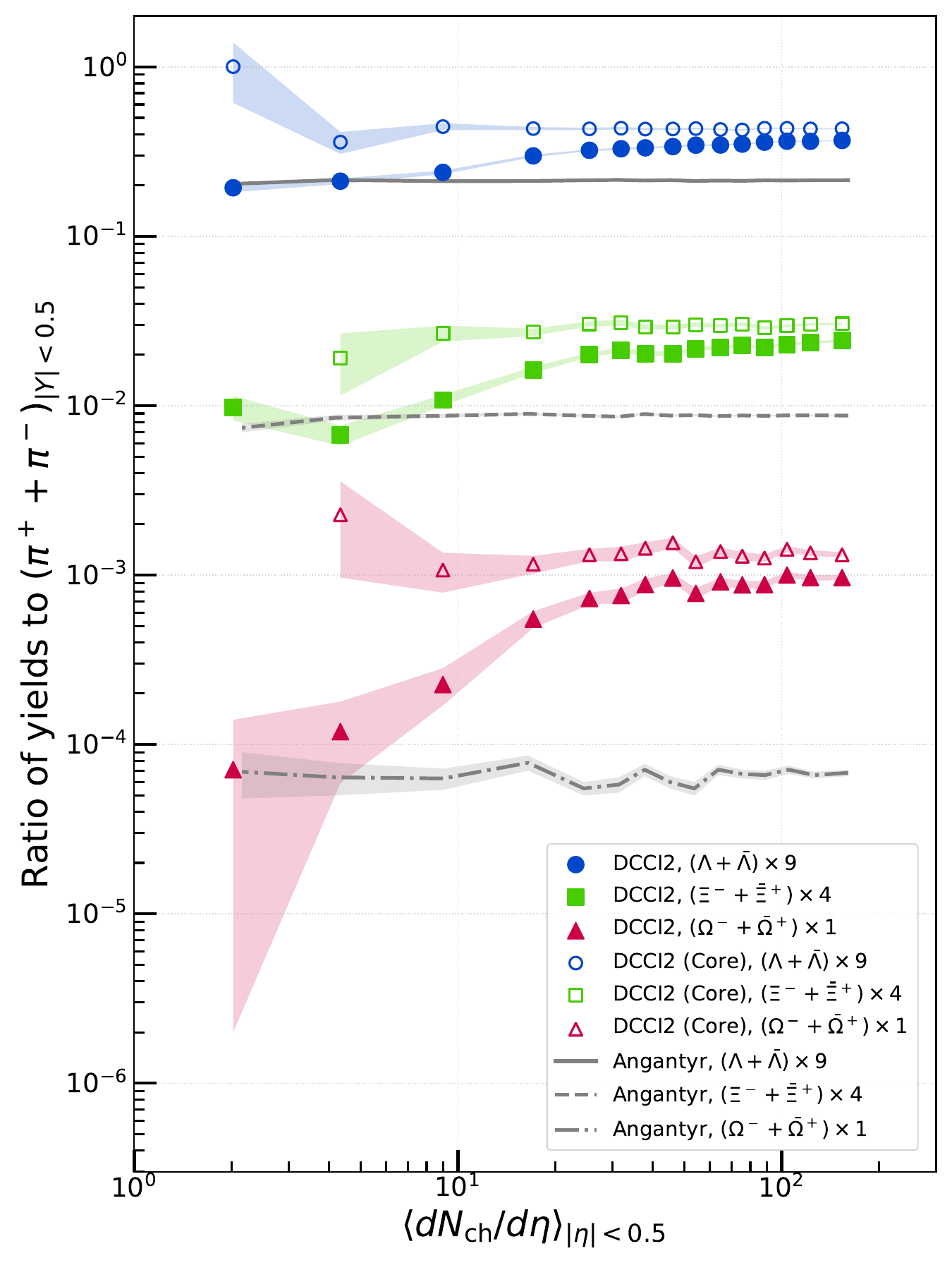}
    \caption{
    Particle yield ratios of $\Lambda + \bar{\Lambda}$ (blue circles), $\Xi^- + \bar{\Xi}^+$ (green squares), and $\Omega^- + \bar{\Omega}^+$ (red triangles) to charged pions at midrapidity as functions of $\langle dN_{\mathrm{ch}}/d\eta \rangle_{|\eta|<0.5}$ obtained from the DCCI2 and PYTHIA8 Angantyr models.
    Solid symbols denote the full DCCI2 results, while open symbols denote the results from the core component only.
    For comparison, the lines represent the results from the Angantyr model for each particle species.
    The shaded areas in all plots and lines indicate the statistical uncertainties.
    The centrality classes are defined in 5\% intervals for the range of 0--50\% and 10\% intervals for 50--100\%.
    }

    \label{Fig:Strangeness Enhancement}
\end{figure}

Figure~\ref{Fig:Strangeness Enhancement} shows the particle yield ratios of strange baryons ($\Lambda+\bar{\Lambda}$,  $\Xi^{-}+\bar{\Xi}^+$, and $\Omega^- + \bar{\Omega}^{+}$) to charged pions ($\pi^- + \pi^+$) as functions of the charged-particle multiplicity at midrapidity in $\mathrm{O}+\mathrm{O}$ collisions calculated within the DCCI2 model. 
The yield ratios of strange baryons (hyperons) to charged pions are obtained as
\begin{equation}
    R=\frac{\langle dN_{\text{hyperon}}/dY \rangle_{|Y|<0.5}}{\langle dN_{\text{pion}}/dY \rangle_{|Y|<0.5}}\,.
\end{equation}

A clear monotonic increase of these yield ratios is observed for all strange baryon species, representing the characteristic signature of strangeness enhancement \cite{PhysRevLett.48.1066,KOCH1983151}.
This behavior can be understood in terms of a change in the dominant particle-production mechanism with increasing system activity.
In central collisions, the large energy density achieved in the overlap region favors the formation of a QGP, where strangeness production is enhanced through partonic processes in a thermalized medium. In the DCCI2 framework, this physical picture is realized through the crossover between the core and corona components as discussed in Sec.~\ref{sec:core-corona-fraction}.
In the low-multiplicity (peripheral) region, particle production is dominated by the corona component, and the strangeness ratios are primarily determined by string fragmentation.
As the multiplicity increases toward more central collisions, the fractional contribution of the core component becomes increasingly significant, and the strange-baryon yields reflect the chemical composition of locally equilibrated matter \cite{Ashraf:2024ocb}.
Because the core component produces strange hadrons more efficiently than the corona, the overall yield ratios increase as the system evolves toward a core-dominated regime.
It should be emphasized that these yield ratios never reach those solely from the core components, reflecting that the substantial amount of corona components still remain even in central collisions (See also Fig.~\ref{Fig:CoreCoronaFraction}). 
As a reference, Fig.~\ref{Fig:Strangeness Enhancement} also shows the particle yield ratio from PYTHIA8 Angantyr (with the default parameter set except for $p_{\mathrm{T0Ref}}= 1.67$ GeV to reproduce the centrality dependence of multiplicity at midrapidity in $\mathrm{O}+\mathrm{O}$ collisions), which does not include any influence of equilibrated QCD matter. 
The ratios from PYTHIA8 Angantyr are almost independent of multiplicity, which is one of the manifestations of ``jet universality," namely the string fragmentation being independent of how the string is formed.
The results from the corona components only (not shown) are almost identical to those from PYTHIA8 Angantyr.

Notably, the magnitude of the enhancement becomes more pronounced for hadrons with higher strangeness content, such as the $\Omega$ baryon ($|S|=3$).
This trend is consistent with measurements by the ALICE Collaboration in systems ranging from $p + p$ to $\mathrm{Pb} + \mathrm{Pb}$, including $p + \mathrm{Pb}$ and $\mathrm{Xe} + \mathrm{Xe}$, collisions at the LHC energies  \cite{ALICE:2013xmt,ALICE:2016fzo}.
Within the DCCI2 framework, the observed multiplicity dependence of strangeness enhancement in $\mathrm{O}+\mathrm{O}$ collisions arises naturally from the increasing contribution of equilibrated QCD matter toward central collisions.

\section{SUMMARY}
    In this paper, we have performed a comprehensive investigation of hadron production in $\mathrm{O} + \mathrm{O}$ collisions at $\sqrt{s_{NN}} = 5.36$ TeV using the Dynamical Core--Corona Initialization (DCCI2) framework. By updating the initial state generation module to PYTHIA 8.315, we have established a state-of-the-art baseline for $\mathrm{O} + \mathrm{O}$ collisions, providing a crucial bridge between small ($p + p$) and large ($\mathrm{Pb} + \mathrm{Pb}$) colliding systems. Dynamical decomposition into core fluids and corona particles offers a unique perspective on the onset multiplicity of producing equilibrated matter. The main conclusions of the present paper are as follows:

First, pseudorapidity distributions and centrality dependence of charged hadron yields from the DCCI2 model exhibited excellent agreement with the preliminary experimental data from the CMS Collaboration. We quantified the core--corona crossover point as $\langle dN_{\mathrm{ch}}/d\eta \rangle \approx 20$ at midrapidity. Crucially, our analysis reveals that even in the most central (0--10\%) $\mathrm{O} + \mathrm{O}$ collisions, the nonequilibrated corona component still contributes approximately 30\% of the total hadron yields. This finding strongly suggests that intermediate-sized systems cannot be described solely by pure hydrodynamics, and a dynamical two-component approach is indispensable for the analysis of bulk observables.

Second, the analysis of $p_T$ spectra revealed the interplay between collective flow and string fragmentation. We found that the transition point where corona-dominance takes over shifts toward higher $p_T$ for heavier particle species. This mass-dependent shift is a clear manifestation of the radial flow effect originating from the expanding core fluid, which preferentially boosts heavier particles to higher momentum regions. The fact that the proton distribution is dominated by the core component up to a relatively higher $p_T$ range confirms that the DCCI2 framework effectively captures the collective dynamics of the QGP even in $\mathrm{O} + \mathrm{O}$ systems.

Third, we demonstrated that the DCCI2 model naturally explains the strangeness enhancement in $\mathrm{O} + \mathrm{O}$ collisions as a consequence of the increasing fractional contribution of the thermalized core. The enhancement becomes more pronounced for hadrons with larger strangeness content, such as $\Omega$ baryons, reflecting the high strangeness production efficiency of the equilibrated core matter compared to the corona component.

This study provides a robust baseline for several promising research directions. The integration of PYTHIA 8.315 into the DCCI2 framework allows for systematic investigations into the effects of various nuclear density profiles. By applying this framework to other intermediate systems, such as Neon--Neon ($\mathrm{Ne} + \mathrm{Ne}$) collisions, we aim to explore the influence of specific nuclear structures, such as $\alpha$-clustering and nuclear deformation, on final-state observables. Since the corona component is sensitive to the initial nucleon distributions inside colliding nuclei, the DCCI2 model is uniquely positioned to quantify how these fundamental nuclear properties manifest in high-energy collision dynamics. These future investigations will significantly enhance our understanding of the interface between nuclear structure and the evolution of the QGP.

\section*{Acknowledgement}
    We gratefully acknowledge S.-e.~Fujii, A.~Shigeyoshi,  Y.~Tachibana, and S.~Zhao  for fruitful discussions, and Y.~Kanakubo for generously providing us with the DCCI2 code, which is essential for this study.
The work by T.H. was partly supported by JSPS KAKENHI Grant No. JP23K03395.

\bibliography{TEXs/Reference}

\end{document}